# Hybrid chaos synchronization between a ring and line topologies

Elman Shahverdiev

Azerbaijan National Academy of Sciences Institute of Physics
H. Javid Avenue, 31
Baku- 1143, Azerbaijan
e-mail: e.shahverdiev@physics.science.az

Abstract

Hybrid chaos synchronization between ring-line networks topologies is explored on the example of Ikeda modeling–famous multidisciplinary system. It is established that high quality complete synchronization between constituent lasers is a possibility. Some security implications for the computer network hybrid topology are underlined.

1   Introduction

Synchronization as a control method is of certain importance in the scientific and technological fields, see, e.g. [1-2] and references there-in.

Synchronization between systems can help to achieve higher power lasers. This is especially important for the synchronization between Terahertz sources. In recent years it was established that synchronization between Terahertz sources could be potentially helpful to achieve milli Watt powers. Such powers can be vital in creating adequate powers for practical applications. Synchronization between thousands and thousands Josephson junctions present in high temperature superconductors could be helpful in achieving this goal [3].Additionally such synchronization is of immense use to create mobile, small size, portable, cost effective Terahertz devices. Such devises can be used in remote security screening, including at the airports, in detecting fake painting, plastic land mines, etc. see [3] and references there-in.

For the coupled chaotic systems, many different synchronization states have been studied. Complete or identical synchronization [4-5] was the first to be discovered and is the simplest form of synchronization in chaotic systems. Other types of synchronization include: phase synchronization [6]; lag synchronization [7]; inverse synchronization [8] (some researchers use the term antiphase synchronization [9]); generalized synchronization [10]; dual and dual-cross synchronization [11]; cascades and adaptive synchronization [12]; anticipating synchronization [13], etc.
Synchronization in complex systems is of a certain importance in governing and performance improving point of view [14]. This is based on the fact that a chaotic attractor consists of infinite number of periodic orbits, along which the nonlinear system's performance differs. Choosing the "right" periodic orbit the system's yield can be optimized.

As synchronization is associated with communication, a study of existence and stability conditions for synchronization is of paramount importance between networks and their constituents [1].

While focusing on the positive side of the chaos synchronization, one should not forget about the situations when synchronization between interacting systems could be quite harmful. For example, in epileptic patients synchronization between neurons could be the reason for epilepsy seizures [15]. Anticipating synchronization could be quite helpful for diagnostic purposes, e.g. by anticipating epileptic seizures [13, 16]. Another harmful influence of the synchronization has occurred on the iconic London Millennium bridge on the opening day due to the migration of pedestrian onto the bridge. The bridge has nearly collapsed and immediately was closed to the year-long repair work [17].

This paper explores hybrid chaos synchronization in one of the simplest cases of the network(s) based on the Ikeda system-paradigmatic model of chaotic dynamics in time delay systems. Comprehensive review of networks can be found in [18-19] and references there-in.

The originality and importance of this paper is in the spanning a bridge between chaos synchronization and hybrid computer networks. Synchronization is important in chaos-based communication [1]: At the transmitter a message is masked with chaos, then this combined signal is transmitted to the receiver system. At the receiver due to the chaos synchronization between the transmitter and the receiver chaos is regenerated. Deducting the receiver output from the receiver input one can decode the transmitted message [20]. An extra layer of security could be provided by such an approach to communicating data packets between the computers.

Due to the finite speed of information propagation between the interacting systems, feedback and memory effects, etc, time delay systems are widespread in the science and technology, in the natural world [1]. Hyperchaos systems, especially multiple time delay systems are very attractive from the chaos based communication, see, e.g. [21-22] and references there-in. Apart from this, hyperchaos systems can be used to model space-time processes, see e.g. [23].

In this paper we consider hybrid network topology [24] consisting of a ring of three Ikeda systems [25-26] and line (point-to point) topology. By exploring errors' dynamics existence conditions for the complete hybrid chaos synchronization are derived. We also provide the results of numerical simulations to support the theory. These findings are of certain importance in communication between computers, in obtaining high power lasers.

The organization of the rest of this paper is as follows. In Section 2 we briefly introduce the working model and present the results of the analytical study. Section 3 is dedicated to the numerical simulations of hybrid chaos synchronization between the Ikeda models consisting of ring and line topologies. The conclusions are presented in Section 4.

2  Hybrid ring-line topology

In this work in the hybrid ring–line network topology the nodes are described by the delay differential equations based on Ikeda systems [25-26]. Lines connecting the nodes are called links. The network topology is a connection of nodes and links.

Fig.1 shows the schematic description of the simplest case of the ring cavity.

Figure 1

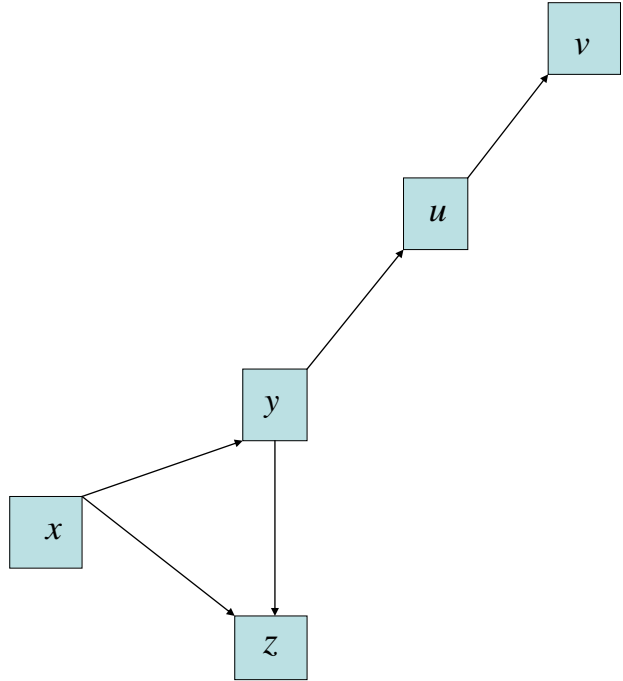

Fig.1. Schematic description of ring-line network topology. Nodes $x$, $y$ and $z$ form the simplest ring network. Nodes $y$, $u$ and $v$ form line network. As underlined above, nodes are Ikeda functional (delay differential) equations: $dx/dt = -\alpha x + m \sin x(t-\tau)$. The definition of the dynamical variables $x$, relaxation rates $\alpha$, feedback rates $m$ and time delay $\tau$ is given below.

    Initially Ikeda model was introduced to describe the dynamics of an optical bi-stable resonator, playing an important role in electronics and physiological studies and is well-known for delay-induced chaotic behavior, see e.g. [25] and references there-in. $x$ is the phase shift experienced by the electrical field in the medium placed in the optical resonator; $\alpha$ is the damping coefficient; $m$ is the field injected into the optical resonator and is proportional to the incident light intensity; delay time $\tau$ is the round-trip time of light in the ring optical resonator.

Later it was established that the Ikeda model or its modifications can be used to describe the dynamics of an opto-electronical, electro-optical and acousto-optical systems and even the dynamics of the wavelength of lasers [26]. In overall Ikeda model is adequate for describing: The propagation of laser intensity in a optical resonator; wavelength chaos; for some solid state lasers, low pressure $CO_2$ lasers (the so-called B class lasers) [26-28]; opto-electronic systems, where $x$ is the voltage fed to the KDP modulator; autoimmune diseases; neuronal models, etc.

Below we consider the hybrid network topology case based on Ikeda modeling:

$$dx/dt = -\alpha x + m_1 \sin x(t-\tau) + m_6 \sin z(t-\tau), \quad (1)$$

$$dy/dt = -\alpha y + m_2 \sin y(t-\tau) + m_7 \sin x(t-\tau), \quad (2)$$

$$dz/dt = -\alpha z + m_3 \sin z(t-\tau) + m_8 \sin y(t-\tau), \quad (3)$$

$$du/dt = -\alpha u + m_4 \sin u(t-\tau) + m_9 \sin y(t-\tau), \quad (4)$$

$$dv/dt = -\alpha v + m_5 \sin v(t-\tau) + m_{10} \sin u(t-\tau). \quad (5)$$

Where $m_1, m_2, m_3, m_4, m_5$ are the feedback strengths; $m_6, m_7, m_8, m_9, m_{10}$ are the coupling strengths between Ikeda systems.
We will take the feedback time delays and connection delay times the same. We also consider the case of identical relaxation coefficients.
By studying errors' dynamics between Ikeda systems one obtains the existence conditions for the complete synchronization:
$$m_1 = m_2 = m_3 = m_4 = m_5 = m_6 = m_7 = m_8 = m_9 = m_{10} \quad (6)$$
These conditions are also known as necessary conditions for the complete synchronization. Unfortunately it is rather difficult to estimate the stability conditions analytically. Further in this work we conduct extensive numerical simulations.

3 Numerical simulations and discussions

This section numerically demonstrates that how the analytical estimates of the previous Section is validated. We simulate the numerical calculations with the help of MATLAB 2008b. Synchronization quality is characterized by the cross-correlation coefficient $C$ between the dynamical variables $x$ and $y$. This coefficient indicates the quality of synchronization: $C = 1$ means perfect synchronization.
We simulate Eqs. (1-5) for the following set of parameters $\alpha = 2, \tau = 3, m_i = 5$, where $i$ runs from 1 to 10.

Initial states are: $x(0) = 2, y(0) = 2.1, z(0) = 2.2, u(0) = 2.3, v(0) = 2.4$.
Fig.2 depicts dynamics of variable $x$.

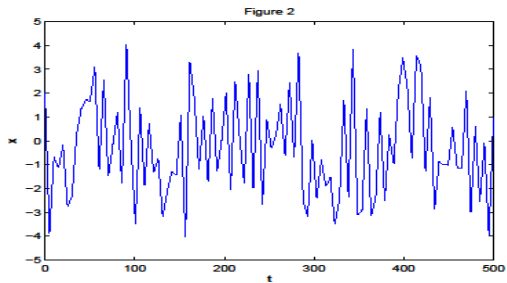

Fig.2 Dynamics of variable $x$ for the set of parameters $\alpha = 2, \tau = 3, m_i = 5$, where $i$ runs from 1 to 10. Dimensionless units.

Fig. 3 demonstrates error dynamics $x - y$ with time. After short transients error $x - y$ approaches zero.

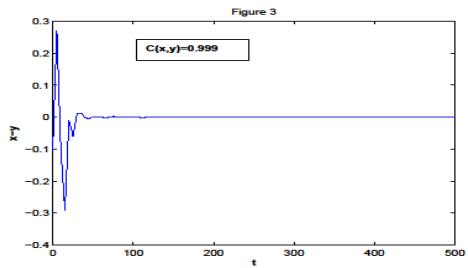

Fig.3 Error $x - y$ dynamics for parameters as for Fig.2. Correlation coefficient between variables $x$ and $y$ is C=0.999. Dimensionless units.

Fig.4 presents dependence between variables $x$ and $z$. This dependence, in full agreement with the theory and numerical simulations is linear: $x = z$

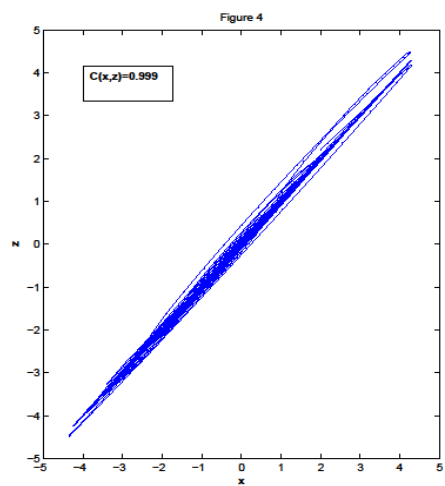

Fig.4 Dynamical variable $z$ versus dynamical variable $x$, for parameters as in Fig.3. Correlation coefficient between variables $x$ and $z$ is C=0.999. Dimensionless units

Figure 5 shows the time dependence of the error $y - v$. In spite of the different initial states $y$ approaches $v$.

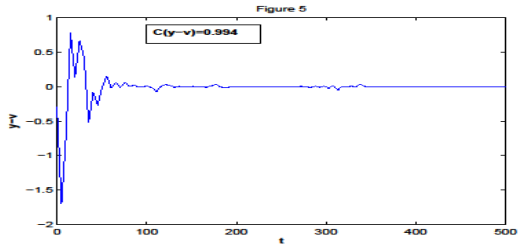

Fig.5 Error dynamics $y-v$ for the set of parameters as in Fig.3. Correlation coefficient between variables $y$ and $v$ is C=0.994. Dimensionless units.

As regards for the numerical simulations, it is worth noting that the initial states for dynamical variables were chosen different. Otherwise the synchronization can be sporadic, accidental.
Below for completeness we also present correlation coefficients between all the nodes.

Table I Correlation coefficients between all the nodes of hybrid network topology.

| C(x, y)=0.999 | C(x, z)=0.999 | C(x, u)=0.999 | C(x, v)=0.995 | C(y, z)=0.999 |
|---|---|---|---|---|
| C(y, u)=0.999 | C(y, v)=0.994 | C(z, u)=0.999 | C(z, v)=0.994 | C(u, v)=0.995 |

4 Conclusions

In this paper we spanned the bridge between chaos control methods and widely used in computer network(s) the hybrid topology. This is the important approach to connect chaos synchronization and computer network(s) topology. As mentioned earlier synchronization is vital in chaos-based communication system to decode the transmitted message [1,20]. Chaos based communication approach could provide an extra layer of security in data exchange between communicating computers.

We also dwell on the advantages and disadvantages of the hybrid network topology in the computer network(s) architecture [24]. Among advantages of hybrid networks topology one should mention its reliability: fault detection and troubleshooting is not difficult. Scalability of hybrid network topology is also deserves attention: it is easy to increase the size of network without disturbing existing architecture. Another advantage of the hybrid network topology is its flexibility: it can be designed according to the requirements of the organization. As hybrid topology is the combination of two or more topologies its effectiveness could be increased by maximizing the strengths of constituent networks or neutralizing the drawbacks of the subunits. Disadvantages of the hybrid network topology are due to the complexity of design and costly hubs (used to connect different networks) and infrastructure. Studied in this paper hybrid network topology is used in the local area networks, wide area networks, etc.

The speed of communication is another important issue in the data packet exchange in in the computer networks. Synchronous optical fiber networks, which could provide communication speed around 40Gb/s. In Terahertz communications [29] the effective data exchange rates exceeding 1Tbit/s (usually on an optical carrier) can be realized. Additionally, communication with a Terahertz carrier wave [29] is also possible. In the case of optical cables between the networks data rates could be even higher than 1Tbit/s. This is due to the fact that wave division multiplexing approach [30] can be exploited. In this approach data packets can be sent using multiple wavelengths.

The studied in this paper configuration could serve as a building block for much more complex network(s) and computer architecture.

# References

1 E.Schoell, H.G.Schuster, (Eds.), Handbook of chaos control, second edition, Wiley-VCH,Weinheim, 2008. https://doi.org/10.1002/9783527622313.

2 E.M. Shahverdiev, Modulated time delays, synchronized Joshepson junctions in high-temperature superconductors and chaotic terahertz waves. Journal of Superconductivity and Novel Magnetism 34 (2021)1125-1132. https://doi.org/10.1007/s10948-021-05837-7.

3 S. Nishijima,.S.Eckroad, A.Marian, et al. Superconductivity and the environment: a Roadmap. Superconductor Science and Technology 26 (2013) 22-24. https://doi.org/10.1088/0953-2048/26/11/113001.

4 L.M. Pecora, T.L.Carroll,Synchronization in chaotic systems. Phys.Rev.Lett. 64 (1990) 821-824. https://doi.org/10.1103/PhysRevLett.64.821.

5 E.M.Shahverdiev, Boundedness of dynamical systems and chaos synchronization, Phys.Rev. E 60 (1999) 3905-3909. https://doi.org/10.1103/PhysRevE.60.3905

6 M. Rosenblum, A.Pikovsky, J.Kurths, From phase to lag synchronization in coupled chaotic oscillators, Phys. Rev. Lett.78(1997) 4193-4196. https://doi.org/10.1103/PhysRevLett.78.4193.

7 E.M.Shahverdiev, S. Sivaprakasam, K.A.Shore, Lag synchronization in time –delayed systems, Phys.Lett.A 292 (2002) 320-324. https://doi.org/10.1016/S0375-9601(01)00824-6.

8 E.M.Shahverdiev, R.A.Nuriev, L.H. Hashimova, et al. Complete inverse synchronization, parameter mismatches and generalized synchronization in the multi- feedback Ikeda model, Chaos , Solitons and Fractals 36 (2008) 211-216. https://doi.org/10.1016/j.chaos.2006.06.026.

9 S.Ghosh, G.K.Sar, S.Majhi, D.Ghosh, Antiphase synchronization in a population of swarmalators, Phys. Rev. E 108 (2023) 034217. https://doi.org/10.1103/PhysRevE.108.034217.

10 E.M. Shahverdiev, K.A.Shore, Generalized synchronization in time delayed systems. Phys.Rev.E 71 (2005) 016201. https://doi.org/10.1103/PhysRevE.71.016201

11  E.M.Shahverdiev, S.Sivaprakasam, K.A. Shore, Dual and dual-cross synchronization in chaotic systems, Optics Communications 216 (2003)179-183. https://doi.org/10.1016/S0030-4018(02)02286-1.

12  E.M.Shahverdiev, P.A.Bayramov, K.A.Shore, Cascaded and adaptive chaos synchronization in multiple time-delay laser systems, Chaos, Solitons and Fractals 42(2009)180-186. https://doi.org/10.1016/j.chaos.2008.11.004

13  Sivaprakasam,S., Shahverdiev,E.M., Spencer,P.S.,and Shore, K.A. :Experimental demonstration of anticipating synchronization in chaotic semiconductor lasers with optical feedback. Phys. Rev. Lett. 87 (2001) 154101. https://doi.org/10.1103/PhysRevLett.87.154101

14  E.Ott, E,M.Spano, Controlling chaos, Physics Today 48, (1995)34-40. https://doi.org/10.1063/1.881461

15  W.L.Ditto, K.Showalter, Introduction: Control and synchronization of chaos, Chaos 7 (1997) 509-511. https://doi.org/10.1063/1.166276.

16  H.U.Voss, Anticipating chaotic synchronization, Phys. Rev. E 61 (2005) 5115-5119. https://doi.org/10.1103/PhysRevE.61.5115.

17  B.Eckhardt, E.Ott,S.H. Strogatz, D.M.Abrams, A. McRobie, Modelling walker synchronization on the Millennium Bridge. Phys.Rev.E 75 (2007) 021110. https://doi.org/10.1103/PhysRevE.75.021110.

18  D.Y.Kenett, M.Perc, S.Boccaletti, Networks of networks-An Introduction, Chaos, Solitons and Fractals 80(2015)1-6.https://doi.org/10.1016/j.chaos.2015.03.016.

19  S.Boccaletti, P.De Lellis, C.I. del Genio, K.Alfaro Bittner,T.Criado, S.Jalan, M.Romance,The Structure and dynamics of networks with higher order interactions,Physics Reports 1018 (2023) 1-64. https://doi.org/10.1016/j.physrep.2023.04.002.

20  G.Perez, H.A.Cerdeira, Extracting Messages Masked by Chaos, Phys. Rev. Lett. 74 (1995) 1970-1973. https://doi.org/10.1103/PhysRevLett.74.1970.

21  E.M.Shahverdiev, K.A.Shore, Chaos synchronization regimes in multiple time delay semiconductor lasers, Phys.Rev.E 77 (2008) 057201. https://doi.org/10.1103/PhysRevE.77.057201.

22  E.M. Shahverdiev, Synchronization in systems with multiple time delays. Phys.Rev.E 70 (2004) 067202 .https://doi.org/10.1103/PhysRevE.70.067202.

23  S.Yanchuk, G. Giacomelli, Spatio-temporal phenomena in complex systems
     with  time delays  J. of Phys. A: Mathematical and Theoretical  30 (2017) 103001.
     https://doi.org/10.1088/1751-8121/50/10/103001.

24 en.wikipedia.org/wiki/Hybrid network/ (accessed 15 February,2025).

25 K. Ikeda, D.Daido. O.Akimoto, Optical turbulence: Chaotic  behavior
    of  transmitted light from a ring cavity. Phys. Rev. Lett. 45 (1980) 709-712.
    https://doi.org/10.1103/PhysRevLett.45.709.

26 T.Erneux, L. Larger, M.W. Lee , J.P. Goedgebuer, Ikeda Hopf  bifurcation
     revisited, Physica D 194(2004) 49-64. https://doi.org/10.1016/j.physd.2004.01.03

27 Y.L. Khanin, Low frequency dynamics of lasers, Chaos 6(1996) 373-380.
      https://doi.org/10.1063/1.166181.

28 F.T. Arecchi, G.Giacomelli, A.Lapucci, et al. Dynamics of a $CO_2$
    with delayed feedback: The short delay regime. Phys. Rev. A 43(1991)
    4997- 5004. https://doi.org/10.1103/PhysRevA.43.4997.

29 M.F. Fitch, R. Osiander, Terahertz waves for communications and sensing.
   Johns Hopkins APL Technical digest 25 (2004) 348-355.

30 A. Argyris, D.Syvridis, L. Larger, et al. Chaos- based  communications at high
    bit  rates using commercial  fibre-optic links. Nature 438 (2005) 343-346.
    https://doi.org/10.1038/nature04275.

.

Figure captions

Fig.1 Schematic description of the simplest case of the ring-line network topologies. See the text for more details.

Fig.2 Dynamics of variable $x$ for the set of parameters $\alpha = 2, \tau = 3, m_i = 5$, where $i$ runs Fig from 1 to 10. Dimensionless units.

Fig.3 Error $x - y$ dynamics for parameters as for Fig.2. Correlation coefficient between variables $x$ and $y$ is C=0.999. Dimensionless units.

Fig.4 Dynamical variable $z$ versus dynamical variable $x$, for parameters as in Fig.3. Correlation coefficient between variables $x$ and $z$ is C=0.999. Dimensionless units.

Fig 5 Error dynamics $y - v$ for the set of parameters as in Fig.3. Correlation coefficient between variables $y$ and $v$ is C=0.994. Dimensionless units.